\documentclass[
    ,final            
  ]
  {aipproc}
\usepackage{graphicx,psfrag,bbm,bm,dcolumn,color,mathrsfs,latexsym,amsfonts,amssymb}

\layoutstyle{6x9}

\begin{document}

\title{On the statistics of quantum expectations 
\\for systems in thermal equilibrium\footnote{Dedicated to Giancarlo Ghirardi
on the occasion of his 70th birthday.}}

\classification{05.30.Ch, 05.30-d, 03.65.-w}
\keywords      {quantum statistical mechanics, quantum ensemble theory, coherent states}

\author{Giovanni Jona-Lasinio}{
  address={Dipartimento di Fisica, Universit\`a di Roma ``La Sapienza'',\\ Piazzale A. Moro 2, Roma 00185, Italy}
,altaddress={Istituto Nazionale di Fisica Nucleare, Sezione di Roma 1, Roma 00185, Italy}
}

\author{Carlo Presilla}{
  address={Dipartimento di Fisica, Universit\`a di Roma ``La Sapienza'',\\ Piazzale A. Moro 2, Roma 00185, Italy}
,altaddress={Center for Statistical Mechanics and Complexity, Istituto Nazionale per la Fisica della Materia, Unit\`a di Roma 1, Roma 00185, Italy}
}

\begin{abstract}
The recent remarkable developments in quantum optics, 
mesoscopic and cold atom physics have given reality to wave functions.
It is then interesting to explore the consequences of assuming 
ensembles over the wave functions simply related to the canonical 
density matrix. 
In this note we analyze a previously introduced distribution over wave 
functions which naturally arises considering the Schr\"odinger 
equation as an infinite dimensional dynamical system.
In particular, we discuss the low temperature fluctuations of the 
quantum expectations of coordinates and momenta for a particle in a
double well potential.
Our results may be of interest in the study of chiral molecules.  
\end{abstract}

\maketitle


\section{Introduction}
Quantum statistical mechanics is usually based on the canonical ensemble 
described by the density matrix 
$\bm{\rho}_\beta = Z^{-1} \exp(-\beta \mathbf{H})$, 
with $Z=\mathrm{tr} \exp(-\beta\mathbf{H})$.
In his celebrated book on statistical thermodynamics, Schr\"odinger~\cite{S}
remarked that ``this assumption is irreconcilable with the very foundations of
quantum mechanics'' because microsystems in general will not be in an energy
eigenstate. 
The reason why this assumption can be accepted is that a more consistent 
attitude ``leads to the same thermodynamical results''.
The more consistent attitude, according to Schr\"odinger, consists in 
introducing an ensemble over the states (wave functions) in which one 
attributes the same probability to all the highly degenerate energy 
states of a macroscopic system.
He then shows that for a subsystem one expects the canonical ensemble 
to be approximately correct.
In a recent paper \cite{GLTZ} Schr\"odinger's statement has been substantially
strengthened. See also \cite{Bloch}.

In the last decade the remarkable developments in quantum optics, 
mesoscopic and cold atom physics have given reality to wave functions.
It is enough to think of the many realizations of  ``Schr\"odinger cats''
or of coherent states.
It is then interesting to explore the consequences of assuming 
ensembles over the wave functions simply related to the canonical 
density matrix. 
A guide in the choice of an ensemble is the observation that the 
Schr\"odinger equation, $i\partial_t \psi = H \psi$, and its 
complex conjugate can be considered as an infinite dimensional 
Hamiltonian system in the variables $\psi$ and $\psi^*$ 
with Hamiltonian $\langle\psi,\mathbf{H}\psi\rangle$.
It is therefore natural to concentrate our attention on measures 
invariant under this dynamics.
In~\cite{Jona} an ensemble over the wave functions was introduced,
\footnote{We became aware recently that the same ensemble was introduced 
also in \cite{BH}. 
While some general motivations of \cite{Jona} and \cite{BH} are
essentially the same, \cite{Jona} dealt with a specific problem 
which is further analysed in the present paper.}
hereafter named Schr\"odinger-Gibbs (SG) ensemble, with formal measure
\begin{equation}
  \label{SGmeasure}
  d\mu_\mathrm{SG} (\psi) = Z_\mathrm{SG}^{-1} 
  e^{-\beta\langle\psi,\mathbf{H}\psi\rangle}
  \delta\left( 1-\langle\psi,\psi\rangle\right)
  \prod d\psi,
\end{equation}
where $\langle \cdot,\cdot \rangle$ is the usual scalar product in 
the Hilbert space.
A simple calculation shows that in terms of (\ref{SGmeasure}) the
canonical ensemble can be written  
\begin{equation}
  \bm{\rho}_\beta = \int | \psi \rangle \langle \psi|
  \left( \sum_k \delta(\psi-\psi_k) \right)
  d\mu_\mathrm{SG}(\psi), 
\end{equation}
where $\psi_k$ are the \emph{unnormalized} eigenstates of $\mathbf{H}$,
assumed to have a discrete spectrum.

In \cite{Jona} the motivation was to calculate the distribution of quantum 
expectation values of the operators $\mathbf{q}$ and $\mathbf{p}$
induced by the SG ensemble.
A formula for such a distribution was found at low temperature and 
the wave functions giving the largest contributions were characterized 
in terms of an appropriate Legendre transform of the ground state energy 
of the system in an external field.
These wave functions in the case of the harmonic oscillator are the
usual coherent states and, by analogy,
we shall adopt this name also in the general case.

The distribution for the expectation values
$\langle\mathbf{q}\rangle$ and $\langle\mathbf{p}\rangle$ 
is different from that obtained with the usual quantum canonical ensemble.
For example, in the latter case for reflection invariant systems
$\langle\mathbf{q}\rangle$ does not fluctuate at all, 
a rather unphysical result. 

In this paper, after recalling the main steps in~\cite{Jona}, we analyze 
in detail the case of a one-dimensional double well, 
which is an ubiquitous system in physics and deeply non classical at low 
energies. 
The result is a Gibbsian distribution of the expectation values
$\langle\mathbf{q}\rangle$ and $\langle\mathbf{p}\rangle$ of the form
$P(\langle\mathbf{q}\rangle,\langle\mathbf{p}\rangle) 
\simeq \exp\{-\beta[\langle\mathbf{p}\rangle^2/(2m)+ 
V_\mathrm{eff}(\langle\mathbf{q}\rangle)]\}$,
where $V_\mathrm{eff}(\langle\mathbf{q}\rangle)$ is an effective potential 
with a single minimum.
This minimum is at zero for the symmetric double well or near the point 
where the ground state function is concentrated in the asymmetric case.
This result may be of interest in connection with pyramidal molecules,
like ammonia or other molecules potentially chiral. 
For these systems which are adequately described, 
as far as the inversion degrees of freedom are concerned, 
by a symmetric double well \cite{CJL,JLPT}, 
fluctuations of $\langle\mathbf{q}\rangle$ correspond to fluctuations 
of the electric dipole moment and could be in principle observable
providing thereby a test of the statistical assumption (\ref{SGmeasure}).

\section{Low temperature limit}
We discuss first the harmonic oscillator defined by the Hamiltonian 
$\mathbf{H} = \mathbf{p}^2/2m + m \omega^2 \mathbf{q}^2/2$.
For this system the Heisenberg equations of motion of the canonical
operators $\mathbf{q}$ and $\mathbf{p}$,
namely $\dot\mathbf{q}=\mathbf{p}/m$ and 
$\dot\mathbf{p}=-m\omega^2 \mathbf{q}$,
bring to the $c$-number equations for the expectations
\begin{eqnarray}
  \label{hfhoq}
  \dot{\langle\mathbf{q}\rangle}&=&\langle\mathbf{p}\rangle/m 
  \\
  \dot{\langle\mathbf{p}\rangle}&=&-m\omega^2\langle\mathbf{q}\rangle. 
  \label{hfhop}
\end{eqnarray}
In the following, we will use the notation 
$\langle\mathbf{q}\rangle=q$ and $\langle\mathbf{p}\rangle=p$. 
The Hamiltonian flow (\ref{hfhoq}-\ref{hfhop}) admits the canonical invariant 
measure 
$Z^{-1} \exp[-\beta H(q,p)] dqdp$, where $H(q,p)=p^2/2m +m\omega^2q^2/2$
and $\beta$ is a constant. 
We want to show that this result can be obtained from the formal SG-measure
(\ref{SGmeasure}).
The probability density of the expectation values $q$ and $p$ 
is given by
\begin{equation}
  \label{Pqp}
  P(q,p) = \int 
  \delta\left( q-\langle\psi,\mathbf{q}\psi\rangle\right)
  \delta\left( p-\langle\psi,\mathbf{p}\psi\rangle\right)
  d\mu_\mathrm{SG}(\psi).
\end{equation}
The Fourier transform of the above expression can be evaluated exactly 
and transforming back to the variables $q$ and $p$ one finds 
\begin{equation}
  \label{Pqpho}
  P(q,p) = \frac{\beta\omega}{2\pi} \exp[-\beta(p^2/2m+m\omega^2q^2/2)].
\end{equation}

It is interesting to determine which kind of wave functions contribute
mostly to $P(q,p)$ in the low temperature regime $\beta \to \infty$.
To this purpose we have to minimize $\langle\psi,\mathbf{H}\psi\rangle$
with the three constraints 
$\langle\psi,\psi\rangle=1$,
$\langle\psi,\mathbf{q}\psi\rangle=q$, and
$\langle\psi,\mathbf{p}\psi\rangle=p$.
Once more the calculation can be performed exactly 
and one finds that the minimizing wave functions are 
\begin{equation}
  \psi_{q,p}(x) = \left( \frac{m\omega}{\pi\hbar}\right)^{1/4}
  \exp\left[\frac{i}{\hbar}px -\frac{1}{2} \frac{m\omega}{\hbar}
    (x-q)^2 \right],
\end{equation}
\textit{i.e.} the harmonic oscillator coherent states.

We wish now to generalize the above results to systems described by 
Hamiltonians $\mathbf{H}=\mathbf{p}^2/2m +V(\mathbf{q})$ 
with non quadratic potentials $V(\mathbf{q})$.
In that case, the density $P(q,p)$ cannot be evaluated exactly,
however, we can estimate it for low temperatures.
In fact, for $\beta$ large from steepest descent we have
\begin{eqnarray}
  P(q,p)&\simeq& \exp[-\beta \langle\psi_{q,p},\mathbf{H}\psi_{q,p}\rangle],
\end{eqnarray}
where $\simeq$ means asymptotic logarithmic equality and
$\psi_{q,p}$ is a normalized state which minimizes the expectation 
of $\mathbf{H}$ with the constraints 
$\langle\psi_{q,p},\mathbf{q}\psi_{q,p}\rangle=q$ and
$\langle\psi_{q,p},\mathbf{p}\psi_{q,p}\rangle=p$.
First we get rid of the latter constraint by putting
\begin{equation}
  \psi_{q,p}(x) = \exp(ipx/\hbar) \phi_q(x)  
\end{equation} 
with $\langle\phi_q,\phi_q\rangle=1$.
The state $\phi_q$ is determined as the ground state $\phi_{\lambda(q)}^0$ 
of the eigenvalue problem 
\begin{eqnarray}
  \label{eigprob}
  \left(
  \mathbf{H}+\lambda\mathbf{q}
  \right)\phi_{\lambda}
  =E_{\lambda}\phi_{\lambda},
\end{eqnarray}
solved self-consistently with the Lagrange multiplier 
$\lambda=\lambda(q)$ specified by the condition
\begin{equation}
  \label{lc}
  \langle\phi_{\lambda}^0,\mathbf{q}\phi_{\lambda}^0\rangle =q. 
\end{equation}
If $E^0_{\lambda(q)}$ is the eigenvalue associated to the ground state 
$\phi_{\lambda(q)}^0$, we have
\begin{equation}
  \langle\psi_{q,p},\mathbf{H}\psi_{q,p}\rangle =
  \frac{p^2}{2m} + \langle\phi_q,\mathbf{H}\phi_q \rangle =
  \frac{p^2}{2m} + E^0_{\lambda(q)} - \lambda(q) q.
\end{equation}
In conclusion, we obtain
\begin{eqnarray}
  P(q,p)&\simeq& \exp\left[-\beta \left(p^2/2m+V_\mathrm{eff}(q)\right)\right],
\end{eqnarray}
where the effective potential
$V_\mathrm{eff}(q)= E^0_{\lambda(q)}-\lambda(q)q$ 
is evaluated by solving the nonlinear eigenvalue problem 
(\ref{eigprob}-\ref{lc}).
In analogy to the harmonic oscillator case, we call 
$\psi_{q,p}(x)=\exp(ipx/\hbar) \phi^{0}_{\lambda(q)}(x)$ 
coherent states.

On the other hand, in the canonical ensemble
for the probability density of the expectations of $q$ and $p$
we immediately obtain 
\begin{equation}
  P(q,p)= Z^{-1} \sum_k e^{-\beta E_k} \delta(q-q_k) \delta(p),
\end{equation}
where $Z=\sum_k e^{-\beta E_k}$ and
$q_k=\langle\varphi_k,\mathbf{q}\varphi_k\rangle$.
Here, $\varphi_k$ are the normalized eigenstates of $\mathbf{H}$ 
with eigenvalues $E_k$.
In particular, for system invariant under reflection we have 
$P(q,p)= \delta(q) \delta(p)$.

\section{Double well systems}
\begin{figure}
  \includegraphics[width=.7\textwidth,clip]{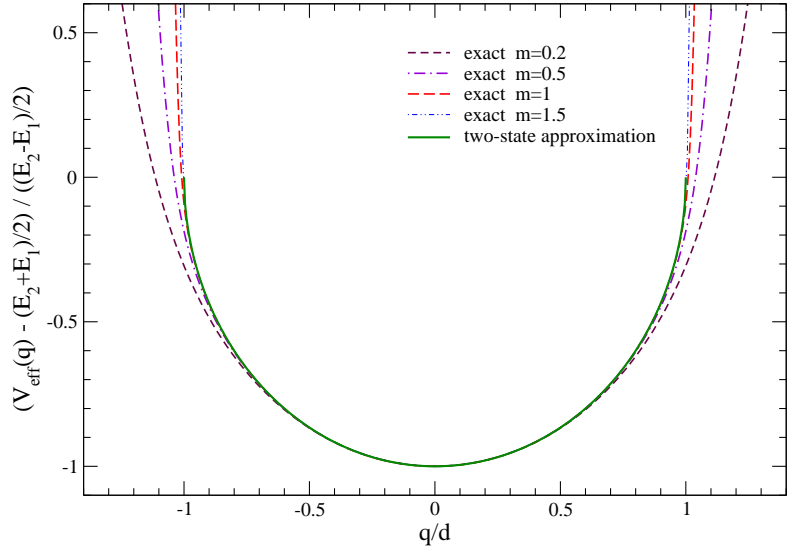}
  \caption{Rescaled effective potential
    $(V_\mathrm{eff}(q) - (E_2+E_1)/2)/((E_2-E_1)/2)$ as a function
    of the rescaled expectation $q/d$ evaluated numerically for the 
    double well potential (\ref{dwp}) with different values of the mass $m$.  
    The solid thick line represents the two-state approximation (\ref{Veffdw}).
    For the other parameters we set, in all cases,
    $\hbar=1$, $x_0=1.5$, and $W_0=1$.
    The values of $E_1$, $E_2$, and $d$ used in the rescaling
    have been evaluated numerically and are: 
    $E_1=3.415753$, $E_2=4.877688$, and $d=1.158335$ for $m=0.2$,
    $E_1=2.582908$, $E_2=2.865508$, and $d=1.268715$ for $m=0.5$,
    $E_1=1.970442$, $E_2=2.012262$, and $d=1.353385$ for $m=1$,
    $E_1=1.64383345$, $E_2=1.65329839$, and $d=1.38670188$ for $m=1.5$.
}
  \label{fig1}
\end{figure}

We consider a particle in a symmetric double well potential of the form
\begin{equation}
  \label{dwp}
  V(\mathbf{q}) = W_0 (\mathbf{q}^2 - x_0^2)^2
\end{equation}
and solve numerically the nonlinear eigenvalue problem 
(\ref{eigprob}-\ref{lc}) to obtain the effective potential $V_\mathrm{eff}(q)$.
This task is accomplished efficiently by the selective relaxation algorithm 
\cite{PT}.
The results are reported in Fig.~\ref{fig1}.
We expect that in this problem one can approximate 
the original Hamiltonian $\mathbf{H}$ with a two-state Hamiltonian 
restricted to the lowest two eigenfunctions of $\mathbf{H}$ 
corresponding to the splitting of the ground state induced by tunneling.
For this reason, we report in Fig.~\ref{fig1} both the exact 
numerical calculations and the two-state approximation.
In the latter case, the effective potential can be evaluated analytically 
and one finds
\begin{equation}
  \label{Veffdw}
  V_\mathrm{eff}(q) = \frac{E_2+E_1}{2} - 
  \frac{E_2-E_1}{2} \sqrt{1-\left(\frac{q}{d}\right)^2},
\end{equation}
where 
\begin{equation}
  d=(\varphi_1, \mathbf{q} \varphi_2)
\end{equation}
and $\varphi_1$ and $\varphi_2$ are the lowest eigenstates of $\mathbf{H}$
with eigenvalues $E_1$ and $E_2$, respectively.
Figure~\ref{fig1} shows that the two-state approximation is rather good
and, in fact, it becomes more and more accurate in the semiclassical limit, 
for example by increasing the value of the mass $m$.
\begin{figure}
  \includegraphics[width=.7\textwidth,clip]{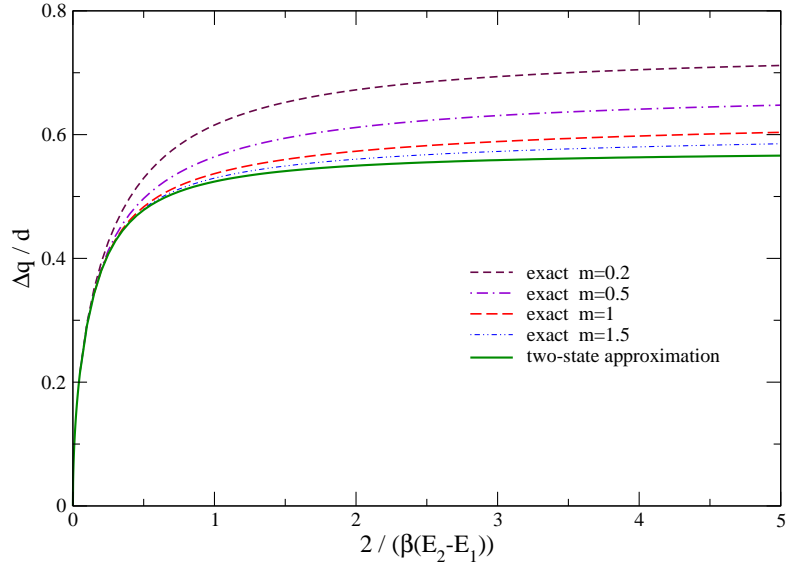}
  \caption{Rescaled fluctuation $\Delta q /d$ as a function
    of the rescaled temperature $2/(\beta(E_2-E_1))$ evaluated numerically 
    for the double well potential (\ref{dwp}) with different values of the 
    mass $m$ as in Fig.~\ref{fig1}.
    The solid thick line is the approximated result obtained from   
    $V_\mathrm{eff}$ given by Eq. (\ref{Veffdw}).
}
  \label{fig2}
\end{figure}

In the two-state approximation the coherent states $\psi_{q,p}$
are given by 
\begin{equation}
  \psi_{q,p} (x) = e^{ipx/\hbar} 
  \left[ c_1(q) \varphi_1(x) + c_2(q) \varphi_2(x) \right],
\end{equation}
where
\begin{eqnarray}
  c_1(q) = \frac{1}{\sqrt{2}} \sqrt{1-q/d}
\qquad
  c_2(q) = \frac{1}{\sqrt{2}} \sqrt{1+q/d}.
\end{eqnarray}

In the semiclassical limit the difference $E_2-E_1$ tends to zero 
exponentially and the effective potential becomes flat between $\pm x_0$.
In this limit $d \to x_0$.
This reflects the fact that the two levels become equiprobable for low
temperatures so that any superposition of the corresponding states has the
same probability.
In the same limit, the dispersion 
$\Delta q = (\overline{q^2} - \overline{q}^2)^{1/2}$
tends to $x_0/\sqrt{3}$ as it is apparent in Fig. \ref{fig2}.


\begin{theacknowledgments}
One of us (G.~J.-L.) wishes to express his gratitude to the organizers
of the Quantum Jumps Conference for the kind invitation.
This research was supported by Italian MIUR under PRIN 2004028108$\_$001.
\end{theacknowledgments}

\bibliographystyle{aipproc}

\end{document}